# BaTiO$_3$ thin films as transitional ferroelectrics with giant dielectric response

*Arnoud S. Everhardt[1]†, Thibaud Denneulin[2,3], Anna Grünebohm[4], Yu-Tsun Shao[5], Petr Ondrejkovic[6], Silang Zhou[1], Neus Domingo[7], Gustau Catalan[7,8], Jiří Hlinka[6], Jian-Min Zuo[5], Sylvia Matzen[9], Beatriz Noheda[1,10]\**

Dr. A. S. Everhardt[1†], Dr. T. Denneulin[2,3], Dr. Anna Grünebohm[4], Dr. Y.-T. Shao[5], Dr. P. Ondrejkovic[6], S.-L. Zhou[1], Dr. N. Domingo[7], Prof. Dr. G. Catalan[7,8], Prof. Dr. J. Hlinka[6], Prof. Dr. J.-M. Zuo[5], Dr. S. Matzen[9], Prof. Dr. B. Noheda[1,10*]

[1] Zernike Institute for Advanced Materials, University of Groningen, The Netherlands.
[2] CEMES-CNRS, F-31055 Toulouse Cedex 4, France.
[3] Ernst Ruska-Centre for Microscopy and Spectroscopy with Electrons (ER-C), Forschungszentrum Jülich, 52425 Jülich, Germany.
[4] ICAMS, Ruhr-Universität Bochum, 44801 Bochum, Germany.
[5] Department of Materials Science and Engineering, University of Illinois, Urbana, Illinois, USA.
[6] Institute of Physics of the Czech Academy of Sciences, Na Slovance 2, 18221 Praha 8, Czech Republic.
[7] Catalan Institute of Nanoscience and Nanotechnology (ICN2), CSIC, Barcelona Institute of Science and Technology Campus, Universitat Autonoma de Barcelona, Bellaterra, 08193 Barcelona, Spain.
[8] ICREA, 08193 Barcelona, Spain.
[9] Center for Nanoscience and Nanotechnology, UMR CNRS - Université Paris-Sud, Université Paris-Saclay, Avenue de la Vauve, 91120 Palaiseau, France.
[10] Groningen Cognitive Systems and Materials Centre (CogniGron), University of Groningen, The Netherlands.
† Presently at: Materials Science Division, Lawrence Berkeley National Laboratory, Berkeley, CA 94720, USA.

\* E-mail: b.noheda@rug.nl

Abstract

Proximity to phase transitions (PTs) is frequently responsible for the largest dielectric susceptibilities in ferroelectrics. The impracticality of using temperature as a control parameter to reach those large responses has motivated the design of solid solutions with phase boundaries between different polar phases at temperatures (typically room temperature) significantly lower than the paraelectric-ferroelectric critical temperature. The flat energy landscapes close to these PTs give rise to polarization rotation under external stimuli, being responsible for the best piezoelectrics so far and a their huge market. But this approach requires complex chemistry to achieve temperature-independent PT boundaries and often involves lead-containing compounds. Here we report that such a bridging state is possible in thin films of chemically simple materials such as BaTiO$_3$. A coexistence of tetragonal, orthorhombic and their bridging low-symmetry phases are shown to be responsible for the continuous vertical polarization rotation, recreating a smear in-transition state and leading to giant temperature-independent dielectric response. These features are distinct from those of single crystals, multi-domain crystals, ceramics or relaxor ferroelectrics, requiring a different description. We believe that other materials can be engineered in a similar way to form a class of ferroelectrics, in which MPB solid solutions are also included, that we propose to coin as transitional ferroelectrics.

Phase transitions (PTs) are among the most interesting and ubiquitous phenomena in nature[1]. In materials science, they are responsible for the technological impact of ferromagnets, ferroelectrics[2], shape-memory alloys or memristors[3]. PTs are associated to desirably large, nonlinear changes in the order parameters (magnetization, polarization, resistance, etc.) and susceptibilities. However, they are often also associated to energy losses, resulting from the cost of nucleating one phase into the other.

Ferroelectrics are an interesting class of materials due to their spontaneous polarization and large responses to external stimuli (electric field or stress), which bestows them with the largest existing capacitances and electromechanical responses. Operating a ferroelectric near its transition temperature $T_C$[4–9] induces flattening of the energy potential and maximizes its dielectric and piezoelectric responses. However, the drawback is poor temperature stability. A way around this problem has been found by engineering phase boundaries between two polar phases via changes of a parameter that is robustly fixed during the lifetime of the device, such as composition. This approach requires careful tuning of the chemistry in order to obtain phase boundaries that are temperature-independent, that is parallel to the temperature axis in the temperature-composition phase diagram (known as morphotropic phase boundaries, MPB). This is the case of PZT or PMN-PT[5], the materials with the best performance. These are still used in most applications despite their lead content, as most alternatives lack the required temperature stability. More recently, compositional gradients have been utilized to achieve large dielectric responses[10], again requiring a careful control of the composition.

Early pioneering work[11–15] has established that the origin of the large piezoelectric and dielectric responses around MPB's is the polarization rotation that takes place in low symmetry monoclinic or triclinic phases present or induced at MPBs. In addition, because the MPB features and the low symmetry phases are a result of the need for elastic matching of the different phases coexisting at the MPB[16], these exceptional properties do not always survive when the material is grown in thin film form, which imposes demanding boundary conditions. In this paper we report a mechanism that leads to huge and temperature-independent dielectric responses and large piezoelectric responses in lead-free ferroelectric films of the classical ferroelectric $BaTiO_3$. In these films, the proximity to different stability minima renders evolving polar domain configurations within the same film, including intermediate bridging phases that support rotation of polarization. These materials cannot be considered ferroelectric crystals and we propose the term transitional ferroelectrics to emphasize their unique responses.

**Effect of boundary conditions on the polar state and switching of $BaTiO_3$**

$BaTiO_3$ thin films with thicknesses between 30 and 300 nm have been grown on $NdScO_3$ substrates with $SrRuO_3$ bottom (and optionally also top) electrodes by Pulsed Laser Deposition. Details can be found in the Methods section and in ref.[17]. Two different domain configurations had been reported to exist in these films close to room temperature[18]. One of the structures is the well-known a/c multi-domain phase, consisting of alternating in-plane (*a*-domain) and out-of-plane (*c*-domain) polarized regions, common in tetragonal thin films; the second structure was described as resembling the 90° in-plane domain configuration ($a_1/a_2$) with an additional weak out-of-plane component. However, the observed contrast in the Piezoelectric Force Microscopy (PFM) images was unusually weak for a strongly polar material, such as $BaTiO_3$, and the details of this complex domain structure and phase diagram are still under debate[17–19].

Dark-field Transmission Electron Microscopy (TEM) (**Figure 1a**) has been performed on these $BaTiO_3$ films to shed light on the details of the domain configurations throughout the film. Due to dynamical diffraction, the variations of contrast in dark-field TEM are related to variations of the ferroelectric polarization along the direction of the selected diffracted beam[20,21]. A strong dependence of the domain orientations both on thickness and on electrical boundary conditions is observed. Using *quasi-symmetric* and thick enough (above 10nm) electrodes, the films develop a 180° domain wall close to the center of the films, parallel to the film/electrode interfaces (hereafter referred to as "horizontal" direction), as evidenced in **Figure 1b** by the strong variation of contrast in the middle of the film and parallel to the interfaces. This wall takes place between c-like domains with opposite polarization direction induced by interfacial dipoles, in the top and bottom parts of the film, respectively, and it is similar to previous reports in ultra-thin $BaTiO_3$ films[22]. In **Figure 1c**, the horizontal dark-field image shows periodic variations of contrast inclined at 45° with respect to the surface, which correspond to an a/c ferro-elastic/electric domain structure. The analysis of the strain fields (see Figure S1 in the Supporting Information) has shown in-plane and out-of-plane strain variations at those 45° inclined domain walls. On the other hand, no additional strain was detected at the horizontal domain wall, confirming that the 180° horizontal domain wall is indeed purely ferroelectric. The overall polarization configuration in the film is shown schematically in **Figure 1d**.

The polarization-voltage (hysteresis) loop measured along the out-of-plane direction of the film (**Figure 2a**) gives a saturation polarization of 25-30 $\mu C/cm^2$, similar to bulk $BaTiO_3$ (which is expected for this nearly zero-strain state[23]). However, the loop shows no remanence and the corresponding switching currents are wide and consist of two switching current peaks in each direction, instead of the one peak in each direction typically found for regular ferroelectric switching. This is consistent with the polarization switching of the two sublayers with up and down *c*-polarization, separated by the 180° domain wall observed in Figure 1. At zero field the two sublayers possess opposing c-polarizations, and under a field of +/- 0.1 V (10 kV/cm, see also Figure S2), the polarization switches to a parallel configuration. During this process, the electromechanical coupling is strong with a measured piezoelectric coefficient (subjected to clamping[24]) with $d_{33}$ = 70 pm/V. This $d_{33}$ value is significantly larger than that of typical BaTiO$_3$ epitaxial films[25] and comparable to that of PbZr$_x$Ti$_{1-x}$O$_3$ epitaxial films around the MPB[26,27] or BaTiO$_3$ bulk single crystals[28]. When the films are grown with *asymmetric* electrodes (a thin SrRuO$_3$ or platinum electrode), the horizontal 180° charged domain wall is moved away from the center of the film, close to the surface (**Figure 1e,f**). Concomitantly, the ferroelectric hysteresis loop also changes drastically (**Figure 2b**). The switching current loop now shows a single switching current peak with an internal bias of 0.15 V (20 kV/cm). In addition, a finite remanent polarization at zero field is now present, corresponding to the imbalance of up and down polarization induced by the *asymmetric* configuration (see also Figure S3). The piezoelectric loop displays the same internal bias as the ferroelectric hysteresis and shows increased d$_{33}$ values of 100 pm/V at the bias, or switching, field, showcasing a highly asymmetric structure. Moreover, the films act as a 'strain diode'[29] as there is a significant difference between the piezoelectric constant at large positive (70 pm/V) and negative (20 pm/V) biases. In this case the effect is caused by the asymmetry of the electrodes, rather than by combining opposing ferroelectric and flexoelectric effects[29].

**Temperature-independent giant dielectric permittivity**

When the polarization is measured in the in-plane direction, a symmetric squared ferroelectric hysteresis loop similar to that of high-quality single crystals is obtained (**Figure 3a**). This loop reflects the high crystalline quality with a remanent polarization of ~12 μC/cm$^2$, in agreement with the amount of *a*-polarization in these a/c structures[17]. Interestingly, these properties are only observed after field cycling, not uncommon for ferroelectric materials (see Figure S4). The in-plane dielectric permittivity and loss value (tan δ) (**Figure 3b**) are also typical for a ferroelectric material of excellent crystal quality, showing sharp switching signatures at the coercive field.

In contrast, the dielectric permittivity along the out-of-plane direction shows giant values, significantly larger than predicted for such films[18], with low losses, as observed in **Figure 3c** (see also Figure S5 for additional details). While the in-plane dielectric permittivity as function of temperature (**Figure 3d**) shows a pronounced divergence (it reaches >25 000), as expected for ferroelectric crystals when approaching the ferroelectric-to-paraelectric phase transition ($T_c \sim 130$ °C)[17] and, thus, denoting a large degree of ordering; the out-of-plane dielectric permittivity is remarkably stable under temperature variations and only significantly changes (decreases) above $T_C$. This out-of-plane dielectric permittivity is comparable to that of BaTiO$_3$ single crystals at room temperature[30]. While in BaTiO$_3$ single crystals, the dielectric permittivity shows a strong temperature dependence, in the present case we show a giant temperature-independent dielectric permittivity, which – to the best of our knowledge – is larger than any of the temperature-independent dielectric permittivities reported for epitaxial thin films[8,31–34].

**Vertical gradients**

For a detailed understanding of the remarkable properties observed for these BaTiO$_3$ thin films along the out-of-plane direction, a closer look into the local structure of the films is needed. The films with dissimilar electrodes (**Figure 4a**) provide the opportunity to explore the local strain state across the film thickness unhindered by the horizontal 180° charged domain walls. A thick 320 nm BaTiO$_3$ film is used to increase the dimensions of the different phases that develop across the film thickness as strain relaxes. **Figure 4b** shows a bright-field TEM image that displays 45° inclined a/c domains. However, the 45° domain wall contrast disappears gradually, towards the bottom interface (**Figure 4c,d**). Local TEM polarization mapping, using scanning convergent beam electron diffraction (SCBED)[35–37], has been performed to understand the crystal symmetries involved (**Figure 5a**). Excellent agreement between the experiments (**Figure 5b,d,e**), and the simulated crystal symmetries (**Figure 5c,f**), evidences distinctly different regions across the films: at the top of the film, polarization vectors along [100]/[00-1] are consistent with tetragonal a/c domains (Figure 5b,c). At the bottom area of the film, the domains have orthorhombic symmetry, with polar vectors along [101]/[10-1] (Figure 5e,f). In the middle of the layer, there exists a complex transition region with reduced symmetry (Figure 5d), which cannot be reproduced in simulations, using either tetragonal or orthorhombic structure models. These observations are in agreement with the strain analysis of Figure 4c,d. Here, dark-field electron holography[38] was used to measure the strain because it provides a large field of view (about 400×600 nm$^2$), which allows us to map the whole film. The map and strain profiles reveal strong strain gradients that increase from the bottom interface to the top surface. The rather homogeneous strain at the bottom of the film (Figure 4c) is consistent with [101]/[10-1] orthorhombic domains, which are indistinguishable from a strain point of view. Even though this film is significantly thicker than those used for the electrical measurements shown in Figure 2, similar multi-region mesostructures, though with lower resolution, are found for thinner films (Figure S6,7), and similar electrical results are found in this thick film (Figure S2).

It is, in fact, not entirely surprising that BaTiO$_3$ is able to display a coexistence of phases within a single material. A similar phenomenon has been observed across a so-called Thermotropic Phase Boundary[39], where the co-existence of orthorhombic and tetragonal phases have been observed, along with their bridging monoclinic phases. To understand better the origin of these different phases, we have performed ab-*initio* effective Hamiltonian calculations as well as phase-field simulations (Figure S8-10). It is found that small changes in misfit strain (~0.01%) or energy (~2 meV/f.u.) are sufficient to stabilize either orthorhombic or tetragonal ferroelectric phases (see Figure S8-10). It is then, not unexpected to observe both types of symmetries within the same film. The need for them to coexist at the nanoscale brings intense stresses that deform the ferroelectric phases into lower symmetries and inhomogeneous structures.

**Transitional ferroelectric enabled by polarization rotation**

The BaTiO$_3$ films reported here have remarkable properties, with wide switching current peaks, no remanent polarization, huge dielectric permittivities and increased piezoelectric constants in the out-of-plane direction. Although some of these features may resemble relaxor materials, in this case, the BaTiO$_3$ films include none of the ingredients that are known to give rise to random fields or random bonds in relaxors. The in-plane measurement direction behaves like expected for a high-quality ferroelectric crystal, with a squared large remanent polarization hysteresis loops and a pronounced dielectric anomaly at the phase transition.

In the case of the *asymmetric* electrodes, along the out-of-plane direction, the strain and TEM analyses show a strain gradient, with low symmetry phases bridging the expected tetragonal and orthorhombic phases, to provide a polarization rotation coupled to the strain gradient. This low symmetry phase denote a flat energy landscape that enables continuous polarization rotation between the tetragonal and orthorhombic phase. The huge dielectric constant of >4000 originates from the easy rotation of the polarization even for small fields (see Supporting Information Note S7) due to the continuous polarization rotation mechanism. The temperature-independence of the dielectric constant could be explained by the coexistence of (bridging) phases, which smears out T$_C$ over all the temperatures. Similar materials with a polarization gradient but without polarization rotation[32,40] or BaTiO$_3$ films without low symmetry phases[23,41] do not show such temperature-independent response.

The *quasi-symmetric* electrode configuration shares these functional properties related to the low symmetry phase, namely wide switching current peaks, increased piezoelectric d$_{33}$ coefficient and huge temperature-independent dielectric permittivity, with the *asymmetric* case, as observed in Figure 2. The internal bias in the *asymmetric* configuration is created by uncompensated dipoles (as easily seen in Figure 5a) to create the asymmetric structure, leading to a strain diode, while in this *quasi-symmetric* configuration, the 180° domain wall in the center compensates those two sets of dipoles with two switching peaks to get an effective zero internal bias and more symmetric structure.

Concluding, this work demonstrates that transitional states, enabled by polarization rotation gradients and engineered by utilizing materials with nearly-degenerate, differently- oriented polar phases, have properties that are distinct of those of single crystals, multi-domain crystals, ceramics or relaxor ferroelectrics. We show that, in BaTiO$_3$ thin films grown on NdScO$_3$ substrates with varying thicknesses of SrRuO$_3$ electrodes, these gradients facilitate an electrical-field induced gradual rotation of the polarization. While the in-plane direction shows standard ferroelectric behavior, the out-of-plane direction shows a flat energy landscape. It manages to achieve a giant dielectric permittivity with a large piezoelectric constant as would be characteristic of MPBs, while managing an exceptional temperature-stability without requiring demanding boundary conditions. This combination of properties enables energy-efficient electromechanical functionalities and represents the dielectric equivalent of a magnetic permalloy - with high permittivity that is largely temperature-independent and with dielectric hard and easy axes. Similar mechanisms could be utilized in the design of other low power ferroelectrics, piezoelectrics, dielectrics or shape-memory alloys, as well as in efficient electro- and magneto-caloric cooling.

**Acknowledgements**
The authors are grateful to U. Bhaskar for preliminary piezoelectric measurements, to C. Magén for preliminary TEM measurements, to G. Agnus for developing oxides patterning processes and to N. Robin, P. Muralt and D. Damjanovic for useful discussions. A.S.E. and B.N. acknowledge financial support from the alumni organization of the University of Groningen, De Aduarderking (Ubbo Emmius Fonds), and from the Zernike Institute for Advanced Materials. T.D. acknowledges the European Metrology Research Programme (EMRP) Project IND54 397 Nanostrain and the European Union's Seventh Framework Programme (FP7/2007-2013)/ERC Grant Agreement No. 320832. T.D. thanks Knut Müller-Caspary for technical help with the STEM experiment. A.G. acknowledges funding by the Deutsche Forschungsgemeinschaft (SPP 1599 GR 4792/1-2 and GR 4792/2-1). Y.T.S. and J.M.Z. acknowledge the financial support by the DOE BES (Grant No. DEFG02-01ER45923). Electron diffraction experiments were carried out at the Center for Microanalysis of Materials at the Frederick Seitz Materials Research Laboratory of University of Illinois at Urbana-Champaign. J.H. and P.O. were supported by the Operational Programme Research, Development and Education (financed by European Structural and Investment Funds and by the Czech Ministry of Education, Youth and Sports), Project No. SOLID21 - CZ.02.1.01/0.0/0.0/16_019/0000760). N.D. and G.C. acknowledge financial support by the Severo Ochoa Excellence programme.

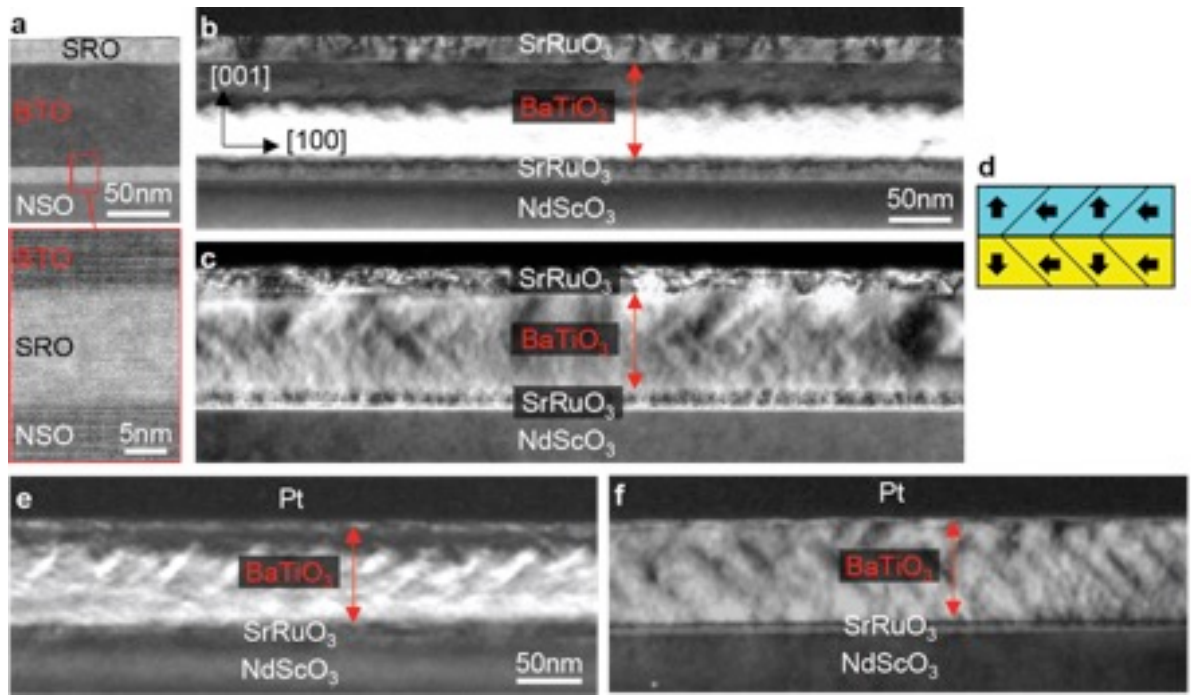

**Figure 1. a,** High resolution STEM-HAADF image of an 80 nm thick BaTiO$_3$ film with a 12 nm SrRuO$_3$ bottom electrode and a 20 nm SrRuO$_3$ top electrode (the *quasi-symmetric* electrode configuration). The bottom image shows a magnified view of the NdScO$_3$/SrRuO$_3$/BaTiO$_3$ interfaces, which shows a defect-free epitaxial growth. **b,** Vertical (002) dark-field TEM images of the previous sample that shows a strong variation of contrast in the middle of the film, attributed to a charged 180° domain wall. **c,** Horizontal (200) dark-field image that shows repeated variations of contrast at 45° to the interfaces, attributed to a/c (or 90°) twin domains. **d,** Simplified schematics of the domain structure and polarization directions. The cyan region represents the averaged up-polarized, and the yellow the down-polarized state, as from the vertical dark-field. **e,** Vertical (002) and **f,** Horizontal (200) dark-field images of an 80 nm thick BaTiO$_3$ film with a (reduced) 6 nm SrRuO$_3$ bottom electrode and a 300 nm thick Pt top electrode (an *asymmetric* electrode configuration).

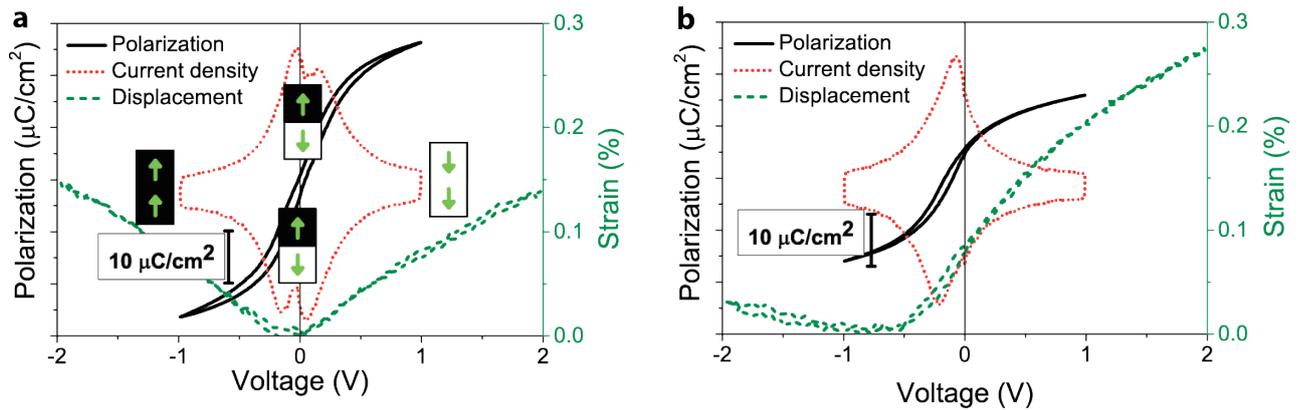

**Figure 2 a,** An 80 nm BaTiO$_3$ thin film with *quasi-symmetric* 12 nm bottom and 20 nm SrRuO$_3$ top electrodes measured at room temperature for a 100 Hz electric field applied along the [001] out-of-plane direction. The insets show a sketch of the local averaged polarization above and below the original 180° horizontal domain wall. **b,** BaTiO$_3$ film with *asymmetric* 6 nm SrRuO$_3$ bottom and 20 nm SrRuO$_3$ top electrodes. Note that the measurement technique can only measure differences in polarization.

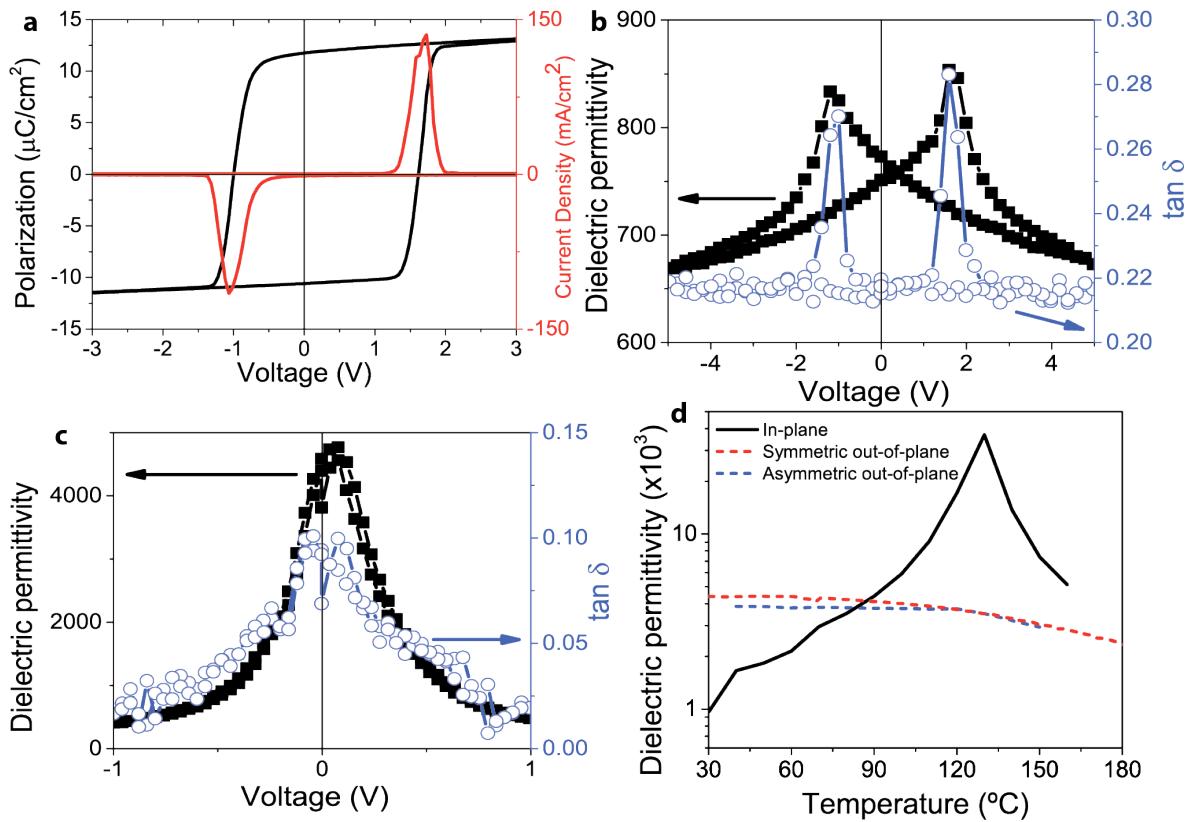

**Figure 3. a,** Ferroelectric hysteresis loop for an 80 nm film at room temperature for a 100 Hz electric field along the [100] in-plane direction with 20 nm SrRuO$_3$ top electrodes and no bottom electrode with 5 μm spacing between the electrodes. Results obtained for this in-plane direction after de-aging for 10$^6$ cycles. **b,** In-plane dielectric permittivity and loss factor (tan δ) along [100] at a field frequency of 1 kHz and a small-signal voltage of 50 mV. **c,** Out-of-plane (so along [001]) dielectric permittivity and loss factor (tan δ) at 1 kHz and a small-

signal voltage of 50 mV using an 80 nm BaTiO$_3$ film with *quasi-symmetric* 12 nm bottom and 20 nm top SrRuO$_3$ electrodes. **d**, Dielectric permittivity as a function of temperature at a DC bias of 0 V at 1 kHz for the in-plane (black) and out-of-plane directions, red for *quasi-symmetric* electrodes and blue for *asymmetric* electrodes. The y-axis is in log(10) scale.

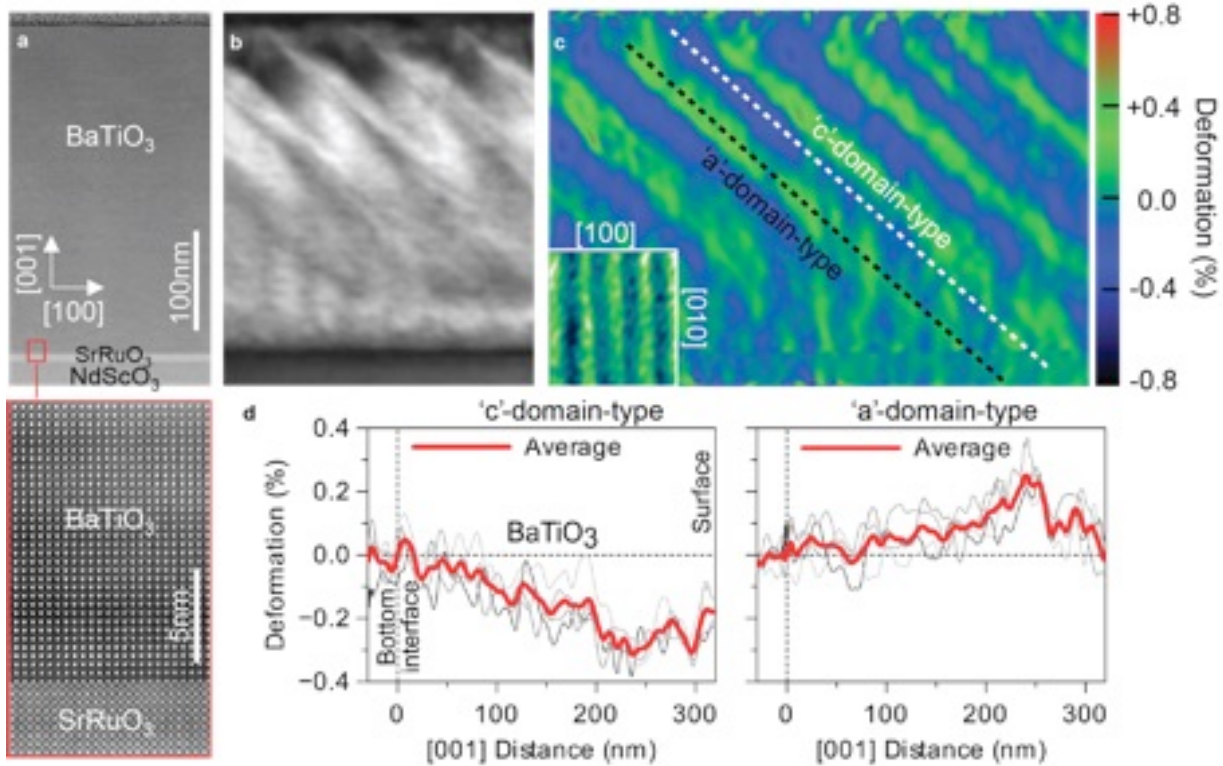

**Figure 4. a**, Aberration-corrected STEM-HAADF (scanning transmission electron microscopy with a high angle annular dark-field detector) image of a 320 nm BaTiO$_3$ film grown on a NdScO$_3$ substrate with a 6 nm SrRuO$_3$ electrode and a thick Pt top electrode. (Top) low magnification image of the film and (bottom) high resolution image of the SrRuO$_3$/BaTiO$_3$ interface, which shows well-defined atomic columns and a defect-free crystal lattice. **b**, Bright-field TEM image of the full film along the [010] direction showing some domain contrast. **c**, In-plane deformation ($E_{xx}$) map obtained by dark-field electron holography, showing (101) domain walls in the BaTiO$_3$ film. $E_{xx}$ is defined with respect to the substrate lattice parameter. The inset in the bottom left corner shows a Piezoelectric Force Microscopy (amplitude) image of the a/c domain structure at the sample surface (viewed along the [001] direction. **d**, Deformation profiles extracted from the dashed regions in (c) running in the [101] direction and plotted as a function of the distance to the bottom electrode (along [001]), at both 'c'-domain-type (left) and 'a'-domain-type (right), as defined by their polarization near the surface. The plots show the profiles of four distinct domains of each type (grey lines), as well as their average (red lines).

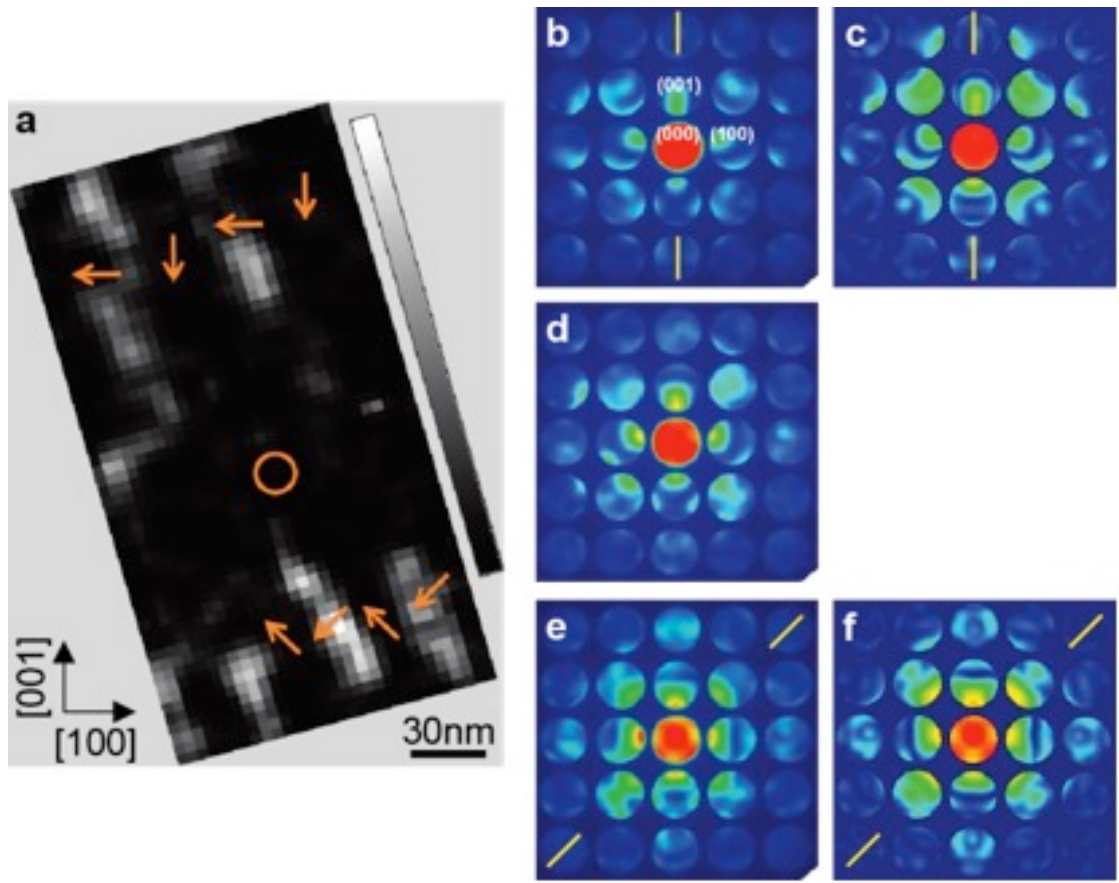

**Figure 5. a**, A 320 nm BaTiO$_3$ film grown on a 6 nm SrRuO$_3$ electrode (not visible in the figure) and a thick Pt top electrode on a NdScO$_3$ substrate, measured along the [010] incidence at room temperature. Image reconstructed from a SCBED dataset where the intensity is determined by the polarity of the domains. The image was obtained by comparing the intensities of four <101> reflections. Orange arrows are a sketch of specific polarization vectors obtained from local CBED patterns. **b-f,** CBED patterns measured or simulated for different regions at the film, showcasing good agreement between measurement and simulation. The orange lines indicate the mirror plane directions, corresponding to the polarization symmetry. **b**, The polarizations at the top surface correspond to tetragonal [100]/[00-1] polarization directions. **c**, Simulated CBED pattern for the tetragonal polarizations. **d**, CBED pattern corresponding to the middle part of the film (this pattern comes from the orange circle in (a)), the polarization transition region. **e,f**, Measured and simulated CBED patterns obtained in the bottom region of the film, corresponding to [101]/[10-1] polarization directions.